\documentclass[sigconf, nonacm]{acmart}

\usepackage{enumitem}

\usepackage{graphicx} 
\usepackage[acronym]{glossaries}
\usepackage{xcolor}
\usepackage{microtype}
\usepackage{xspace}
\usepackage{bm}

\usepackage{tabularx}
\usepackage{booktabs}
\usepackage{array}
\usepackage{svg}

\newcommand{\acrostyle}[1]{\textsc{\textls[10]{#1}}}

\newacronym{csam}{\acrostyle{CSAM}}{Child Sexual Abuse Material\xspace}
\newcommand{\CSAM}{\gls{csam}\xspace}

\newacronym{csem}{\acrostyle{CSEM}}{Child Sexual Exploitation Material\xspace}
\newcommand{\CSEM}{\gls{csem}\xspace}

\newacronym{tou}{ToU}{Terms of Use\xspace}
\newcommand{\TOU}{\gls{tou}\xspace}

\newacronym{gai}{\textls[10]{Gen\kern0,05em AI}}{Generative Artificial Intelligence\xspace}
\newcommand{\GAI}{\gls{gai}\xspace}

\newcommand{\STGB}{\textit{StGB}\xspace}
\newcommand{\figref}[1]{Fig.~\ref{#1}\xspace}
\newcommand{\tabref}[1]{Table~\ref{#1}\xspace}

\newcommand{\law}{\textit{§}}

\newcommand{\wrapemdash}[1]{\textemdash#1\textemdash}

\newcommand{\eg}{e.g.,\xspace}
\newcommand{\ie}{i.e.,\xspace}

\DeclareRobustCommand{\genAIR}{\includegraphics[height=3ex]{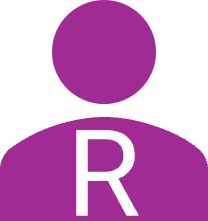}}

\DeclareRobustCommand{\user}{\includegraphics[height=3ex]{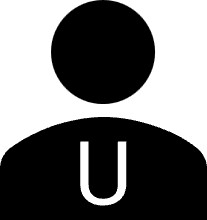}}

\DeclareRobustCommand{\model}{\includegraphics[height=3ex]{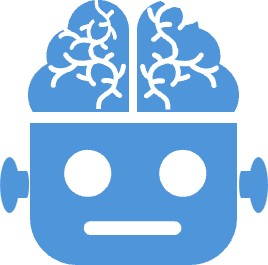}}

\usepackage{tikz}
\usetikzlibrary{decorations.pathreplacing,decorations.pathmorphing}

\usepackage[dvipsnames]{xcolor}
\definecolor{myblue}{RGB}{78,149,217} 
\usepackage{censor}
\usepackage[most]{tcolorbox}

\newcommand{\lawcat}[1]{\tikz[baseline=(X.base)] \node[draw, circle, inner sep=0.5pt](X){\footnotesize {#1}};\xspace}

\makeatletter
\newcommand{\mypara}[1]{%
  \par\addvspace{1ex plus 1ex minus .2ex}%
  \noindent\textbf{#1}\hspace{1em}\ignorespaces%
}
\makeatother

\begin{document}

\title{Criminal Liability of Generative Artificial Intelligence Providers for User-Generated Child Sexual Abuse Material}

\author{Anamaria Mojica-Hanke}
\affiliation{%
  \institution{University of Passau}
  \city{Passau}
  \country{Germany}
  }
\email{mojica01@ads.uni-passau.de}
\orcid{0000-0002-5292-2977} 

\author{Thomas Goger}
\affiliation{%
  \institution{Bavarian Central Office for the Prosecution of Cybercrime}
  \city{Bamberg}
  \country{Germany}
  }
\email{thomas.goger@gensta-ba.bayern.de}
\orcid{0009-0002-9098-8114}

\author{Svenja W\"olfel}
\affiliation{%
  \institution{University of Passau}
  \city{Passau}
  \country{Germany}
  }
\email{svenja.woelfel@uni-passau.de}
\orcid{0009-0007-1789-9103}

\author{Brian Valerius}
\affiliation{%
  \institution{University of Passau}
  \city{Passau}
  \country{Germany}
  }
\email{brian.valerius@uni-passau.de}
\orcid{ 0009-0006-4034-6198}

\author{Steffen Herbold}
\affiliation{%
  \institution{University of Passau}
  \city{Passau}
  \country{Germany}
  }
\email{steffen.herbold@uni-passau.de}
\orcid{0000-0001-9765-2803}

\renewcommand{\shortauthors}{Mojica-Hanke et al.}

\begin{abstract}

The development of more powerful Generative Artificial Intelligence (GenAI) has expanded its capabilities and the variety of outputs. This has introduced significant legal challenges, including gray areas in various legal systems, such as the assessment of criminal liability for those responsible for these models. Therefore, we conducted a multidisciplinary study utilizing the statutory interpretation of relevant German laws, which, in conjunction with scenarios, provides a perspective on the different properties of GenAI in the context of Child Sexual Abuse Material (CSAM) generation. We found that generating CSAM with GenAI may have criminal and legal consequences not only for the user committing the primary offense but also for individuals responsible for the models\textemdash such as independent software developers, researchers, and company representatives. Additionally, the assessment of criminal liability may be affected by contextual and technical factors, including the type of generated image, content moderation policies, and the model’s intended purpose. Based on our findings, we discussed the implications for different roles, as well as the requirements when developing such systems.
\end{abstract}

\ccsdesc[500]{Social and professional topics~Government technology policy}
\ccsdesc[300]{Applied computing~Law}
\ccsdesc[300]{Computing methodologies~Artificial intelligence}

\keywords{Generative AI, Criminal Liability, Governmental Legislation, CSAM}

\begin{teaserfigure}
\centering
  \includegraphics[width=0.9\textwidth]{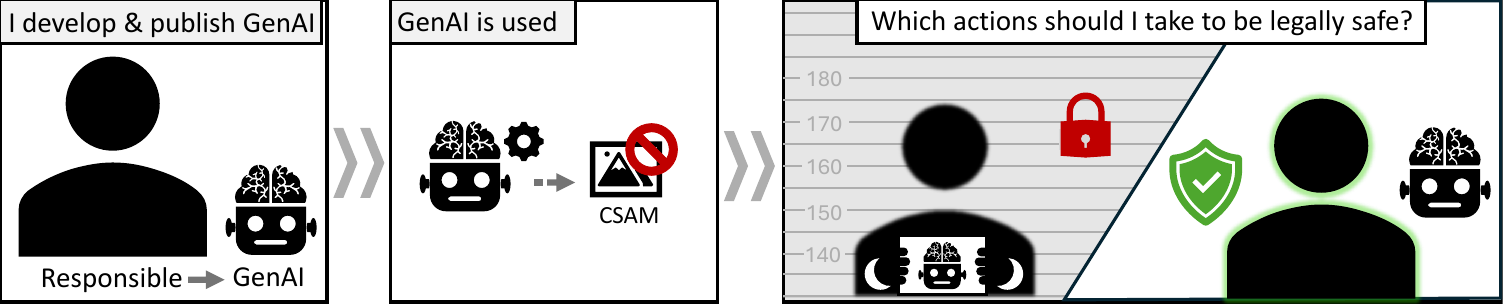}
  \caption{Providing a generative model can expose providers\textemdash \eg developers, researchers, and company representatives\textemdash to legal risk; targeted actions can mitigate it.}
  \Description{A scale that represents the law and divides two sides: being criminal liable or not for developing an AI}
  \label{fig:teaser}
\end{teaserfigure}

\received{20 February 2007}
\received[revised]{12 March 2009}
\received[accepted]{5 June 2009}

\maketitle

\section{Introduction}
Imagine that the CEO of a company developing \GAI, a researcher who publishes their latest model, or a software developer who created \GAI startup is arrested for their work and charged with facilitating the generation and distribution of \CSAM. While this may sound far-fetched, and it is uncomfortable to discuss the negative implications of increasingly powerful AI systems, it is well-known that deepfake technologies,  as well as text-to-image and image-to-image models, have the potential to generate \CSAM content (see, \eg~\cite{netclean2019,  ICMEC2023, hoganlovells, dhs2025impacts, iwf2024aiReport}). Whether the hypothetical scenario presented at the outset is realistic\textemdash that is, whether someone could be arrested for publishing GenAI technology\textemdash is a complex legal question without a simple answer.


As the previous scenario depicts, the rise of \GAI has introduced significant legal challenges, particularly when these systems are misused to produce illegal content such as \CSAM. In fact, the creation of \CSAM with \GAI has opened legal gray areas in several jurisdictions~\cite{inhope2024csam, inhope2024annual, Utrech, CAN, enoughabuse2025stateAIcsam}, particularly regarding the question: 

\begin{description}
\item [\textbf{RQ:}] \textit{Who may be held criminally liable and under what conditions, when \CSAM content is generated using \GAI?} 
\end{description}

Since the answer to this clearly depends on the jurisdiction, we evaluate this scenario based on German criminal law. This choice is based on Germany’s legal and regulatory significance. Not only due to its economic importance as one of the largest economies, but also because of its comprehensive law against \textit{child abuse}. Germany penalizes not only the generation of \CSAM, but also possession and distribution. In addition, Germany is also a party to international agreements to combat \CSAM, such as the \textit{Lanzarote Convention}~\cite{coe2007lanzarote} or the \textit{WeProtect} Global Alliance~\cite{weprotect_global_alliance}. In addition, its legal regime governing the field of \CSAM is heavily influenced by European law, \eg the \textit{Directive 2011/93/EU of the European Parliament and of the Council of 13 December 2011 on combating the sexual abuse and sexual exploitation of children and child pornography}~\cite{euDirective2011_93}.

We answer our research question in this paper with a focus on the criminal liability of the creators and providers of \GAI models, though we also establish the criminal liability of users generating \CSAM. The research method we apply is \textit{statutory interpretation}\textemdash \ie the process interpreting existing laws and regulations\textemdash specifically focusing on relevant laws, \eg \law~\textit{184b–c} German Criminal Code (\STGB). To do so, a set of hypothetical scenarios in which \CSAM is generated using \GAI by a \textit{User} is constructed. These scenarios vary in technical and contextual elements of \GAI usage, such as the location of model deployment and whether a foundational model is involved. This approach allows us to explore how criminal liability may change depending on specific circumstances.

Our analysis reveals that, under current German criminal law: 

\begin{itemize}
    \item Criminal liability for \GAI-generated \CSAM primarily depends on whether the act was committed \textit{deliberately}. The criminal liability will be generally held \textit{individuals} (\ie natural persons), since companies cannot be held criminally liable.
    \item  When (at least conditional) intent is established, the main offender will be the person who generates the imagery, generally the \textit{User} of the \GAI model.
    \item Individuals responsible for the \GAI model\wrapemdash{such as an independent developer, a researcher, or a company responsible (legal representative of a company)} may also be held criminally liable as supporters of the offense.
    \item Whether the aforementioned liability of \GAI providers arises may depend on contextual and technical factors, such as the developers' level of technical expertise, content moderation mechanisms used, or the inclusion of features that facilitate the sharing of generated content, as detailed in this study.
\end{itemize}

The study is organized as follows. In Section \ref{sec::background}, we present the background and the related work in order to: (I) introduce the main concepts that are vital for understanding the scope of what is meant with \CSAM and the subsequent legal analysis; (II) the previous related work about the analysis of generated \CSAM. Then, in Section \ref{sec::methodology}, we describe our approach to answering the research question. Following this, we present the results of applying the method in Section \ref{sec::results}, and discuss them in Section \ref{sec::discussion} in which we provide the implications for developers and \GAI providers. Next, in Section \ref{sec::limitations}, we present the limitations. Finally, we end our study with the conclusions in Section \ref{sec::conclusion}. 

\textit{\textbf{Disclaimer:} This paper provides an academic analysis of criminal liability under German law and is not intended to serve as legal advice.}

\section{Background and Related Work}
\label{sec::background}

Our main target audience is actors involved in the development process of products featuring \GAI, \eg software developers publishing open source software, project managers, and other company representatives who bear legal responsibility and are part of the decision processes, but also researchers who create and publish \GAI models through their work. 

Our goal is to provide this target audience with an understanding of the relevant legal concepts and risks, such that they can make informed decisions about the development of their product. Thus, we position this work as research in the intersection of software engineering, artificial intelligence, and jurisprudence that is focused on enabling decision making that leads to legally safe \GAI software products. Notably, we do not include ethical aspects in our considerations within our work, though we believe such considerations should, and hopefully do, affect the decision-making regarding potentially harmful \GAI system use as well. 

To clarify where our study is situated and what it encompasses for this target audience, we first must establish an understanding of what corresponds or does not as \CSAM, the different styles of \CSAM imagery (Section~\ref{sec::csam}), as well as the relevant aspects of the German legal system (Section~\ref{sec::gLAW}). Additionally, in  Section~\ref{sec::RW}, we present the related work that, to the best of our knowledge, so far has not considered such a focus on understanding possible liability that may be held, especially by the developers and providers of \GAI.

\subsection{Child Sexual Abuse Material (CSAM)}
\label{sec::csam}
\CSAM is a term recognized in numerous legal frameworks and by child-protection organizations to describe any content\wrapemdash{visual, textual, or audio}that depicts minors in sexually abusive situations, \eg~\cite{INTERPOL_2025, Europol_csam, Luxembourg_terminology, US_Department_Justice_2023, Canadian_Child_Protection, ICMEC2023,NCMEC, Inhope_org_2024, euDirective2011_93, IIn_OEA_2021, ACERWC_2021}. The term ``abuse'' makes explicit that minors cannot consent to sexual activity and that such depictions represent a violation of their rights and bodily autonomy.

In addition to the previous considerations, it is important to distinguish \CSAM from a broader category often referred to as \CSEM. While some studies focus on \CSEM, this type of content does not always depict explicit abuse. It may include imagery that may not depict undisguised acts of abuse but still sexualize or exploit minors (\eg ``... a family picture of a young child in a bikini or in her mother's high heels ... being circulated on pornographic websites''~\cite{Luxembourg_terminology}). As such, \CSEM falls outside the scope of our study.

When studying \CSAM imagery, the main category formats are: authentic (real), realistic, photorealistic, and artistic/cartoon-like. The first category, \textit{authentic}, refers to images captured through conventional recording methods (\eg camera) without significant alterations (\eg report~\cite{keech2022cybercrimes}, EU Directive 2011/93~\cite{euDirective2011_93}). The second and third ones, realistic and photorealistic, include synthetically generated or manipulated images that retain the key visual features of genuine scenes (\eg EU Directive 2011/93~\cite{euDirective2011_93}). Photorealistic imagery, in particular, is a subcategory of realistic imagery that aims to closely mimic photography. In contrast, realistic imagery more broadly refers to depictions that appear plausible or lifelike, without necessarily achieving photographic accuracy. Lastly, the fourth category, artistic ones, refers to non-photographic visual depictions\textemdash such as drawings, animations, or computer-generated illustrations that depart from visual realism (\eg~\cite{Grossman2025ai}).

Considering the scope of the previously defined categories and the specific focus of our study—namely, the investigation of generated imagery that depicts child sexual abuse\textemdash we limit our analysis to realistic and photorealistic images. These types of images can be generated using \GAI through text-to-image prompts or based on existing images, which may or may not originate from \CSAM or innocuous imagery~\cite{unicri2024generativeai}. While photorealistic imagery specifically refers to a subset of realistic images that closely resemble actual photographs, note that legal and regulatory frameworks typically do not distinguish between these visual categories. Instead, both are treated under the broader legal classification of realistic representations (\eg~\cite{euDirective2011_93}). Additionally, we do not focus on artistic or cartoon-like depictions, as acts relating to such representations are only partially classified as criminal offenses under German law, which will be further explained in the following sections.

\subsection{Relevant German Legal Aspects}
\label{sec::gLAW}
German law is divided into three main areas: \textit{public law}, which governs the relationship between the state and its citizens; \textit{private law}, which regulates interactions between individuals; and \textit{criminal law}, which defines offenses (\ie acts that break the law and are punishable by the legal system) and penalties. This framework is outlined in key legal texts, including the German Civil Code (\textit{Bürgerliches Gesetzbuch (BGB)}) and the German Criminal Code (\textit{Strafgesetzbuch (StGB)}). Since our research is focused on \CSAM, which is illegal content, the area of our interest is \textit{criminal law}; therefore, most of the laws that are referenced in this study belong to the German Criminal Code, hereinafter referred to as \STGB. 

An important aspect of legal systems is determining who can be held liable for a committed act. In this regard, German \textit{criminal law} stipulates that only a \textit{natural person}, \ie an individual, can be criminally liable. This is because criminal responsibility under the \STGB is based on the premise that a criminal offense requires a conscious decision to break the law, something only an individual can possess, as a \textit{juridical person} (such as a company) cannot possess consciousness. In those cases, the person legally responsible for the company would be the one who may be held liable, and not the employees under its umbrella, \eg developers. 

\begin{figure}[ht!]
  \centering
  \includegraphics[width=0.9\columnwidth]{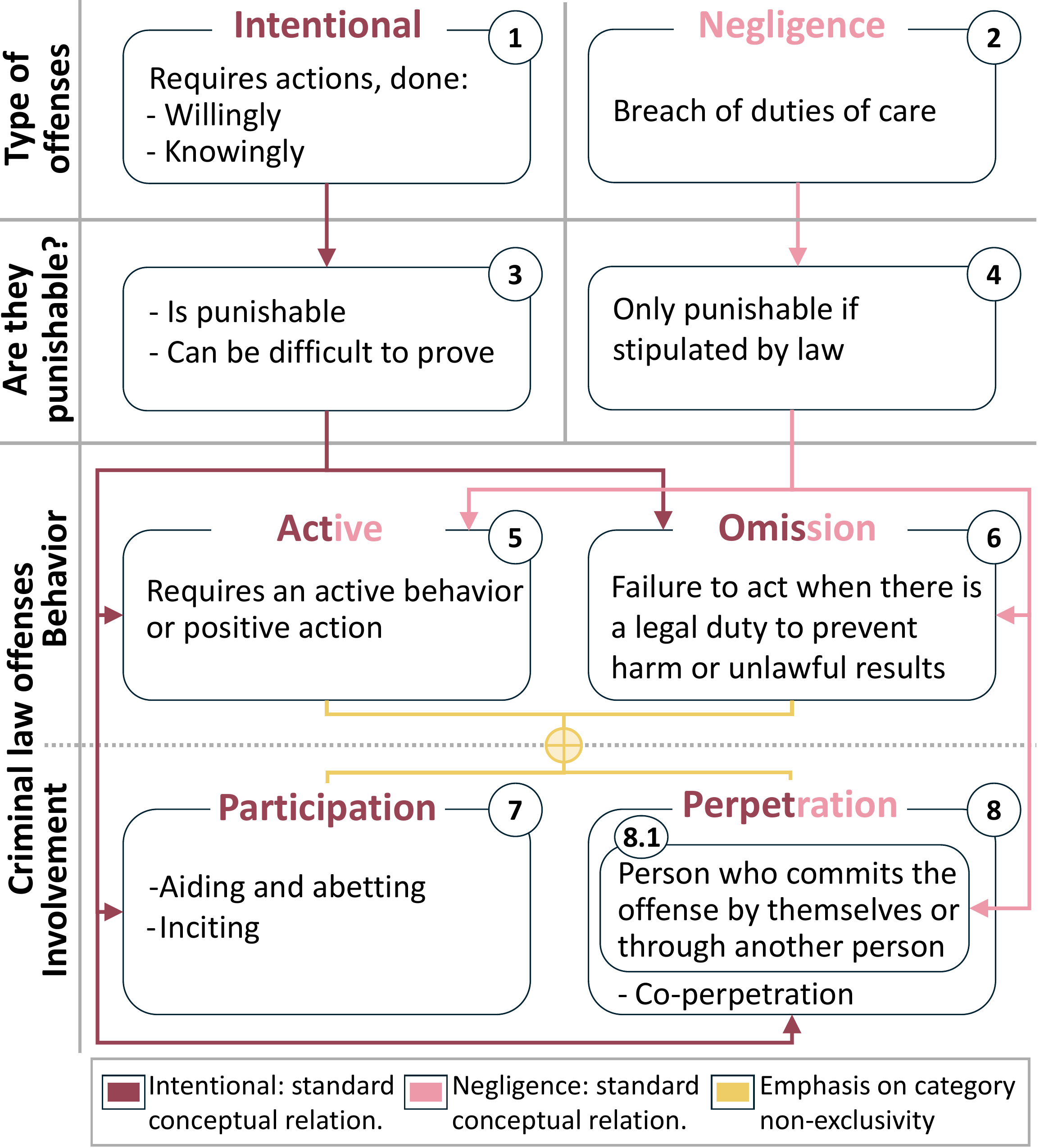} 
  \caption{Relevant aspects of German law. Note that participation only applies to intentional acts, while only perpetration and no co-perpetration exist for the negligent acts.}
  \Description{The image depicts the different concepts of German legislation: negligent and intentional acts; followed by the classification into active and omission, and perpetration and participation.}
  \label{fig:german_law}
  \vspace{-0.3cm}
\end{figure}

Furthermore, \law~\textit{15}~\STGB classifies acts according to the intent behind the committed offense. This categorization divides offenses into two major groups: \textit{intentional} acts and \textit{negligent} acts (see \figref{fig:german_law}). As the name indicates, \lawcat{1}~\textit{intentional} offenses occur when an individual acts \textit{knowingly} or \textit{willingly}. In contrast, \lawcat{2}~\textit{negligent} acts arise from a breach of the duty of care, meaning that harm results from a failure to take the precautions that a reasonable person would take under the circumstances~\cite{kudlich2025beckok, hilgendorf}.

Another important aspect is determining whether an offense is punishable under the \textit{criminal law}. This concept is illustrated in the second row of \figref{fig:german_law}: On one side \lawcat{3}~\textit{intentional acts} are punishable by default, although the authorities must still prove that the act was indeed done deliberately (\ie intentionally); on the other, \lawcat{4}~\textit{negligent acts} are punishable only if the law explicitly stipulates it. Note that the intention behind an act is not always clear. For instance, consider the person who created an image editing program that is later used to produce illegal content. It can be challenging to determine whether they intended their software to be used in an illegal manner or if it was misused by others.

When analyzing \textit{intentional and negligent offenses}, they can be further categorized based on the type of behavior: \lawcat{5}~\textit{active action} and \lawcat{6}~\textit{omission}~\cite{StGB}. The former, as the name indicates, requires an active behavior or positive action; the latter is defined as the failure to act when there is a legal duty to prevent harm or unlawful results. 

Within \textit{intentional} offenses, there is a further categorization based on the type of involvement of the individual: \lawcat{7}~\textit{participation} and \lawcat{8}\textit{perpetration}~\cite{StGB}. Under \law~\textit{25}~\STGB, a \textit{perpetrator} is the person who commits the offense\textemdash either directly by personal action, indirectly by using another as an instrument, or jointly with others in \textit{co-perpetration}.  In contrast, under \law\law~\textit{26, 27}~\STGB, a \textit{participator} is who either instigates (\ie person who persuades or encourages to commit the crime) or assists (\ie aids and abets) the offense. In the case of \textit{negligent} offenses, every responsible person is a \lawcat{8.1}~\textit{perpetrator}, and \textit{co-perpetration} is not possible.

Note that these two different categorizations\wrapemdash{by behavior and by involvement} are orthogonal, not mutually exclusive. In any given intentional offense, whether committed through action or omission, an individual may be either the principal offender (\textit{perpetrator}) or a secondary \textit{participant} (\textit{participator}), as illustrated in \figref{fig:german_law}, excepting the cases previously mentioned for \textit{negligent acts}.  Accordingly, one may bear primary liability as the main offender, or secondary liability for supporting the offense.

In addition to these main concepts of \textit{criminal law}, it is important to note that different areas of the law can intersect or influence one another. For instance, the \textit{public law} can influence the \textit{criminal law}. In particular, public laws may define legal standards and establish regulatory frameworks that determine what is considered lawful or unlawful behavior, or impose legal duties on individuals and organizations. These standards can then serve as reference points when \textit{assessing criminal liability}, especially in cases involving negligence, or harm resulting from non-compliance.  Regarding \GAI, it is \eg possible that by this mechanism propositions of the AI Act will develop implications for the assessment of criminal liability, as we discuss later.

\subsection{Related Work}
\label{sec::RW}
To the best of our knowledge at the time of the publication, no existing studies have addressed a similar legal analysis regarding the criminal liability of providers. Instead, past work focused solely on discussing the legality of \CSAM generation in different jurisdictions~\eg~\cite{iwf2024aiReport, keech2022cybercrimes, inhope2024annual, ICMEC2023, inhope2024csam, NCMEC,dhs2025impacts,CAN,Utrech, kokolaki2025unveiling}.  There are also some other studies that present guidelines that \GAI providers should follow to prevent \CSAM generation~\eg~\cite{iwf2024aiReport, unicri2024generativeai, thorn2024safety, thorn2024solutions, nist2024comments, nist2024synthetic, thiel2023generative, Grossman2025ai, dorotic2023child}. The related work most relevant for this study is the research by \textit{Emmanouela Kokolaki} and \textit{Paraskevi Fragopoulou}~\cite{kokolaki2025unveiling}, and the report by \textit{INHOPE}~\cite{inhope2024csam}, in which they investigate and explicitly mention different laws that apply in multiple legal systems \CSAM,  and define it. However, those studies do not analyze the liability of providers and how different properties of the \GAI may affect the assessment of it.

\section{Method}
\label{sec::methodology}

This study combines statutory interpretation under German law with a scenario-based analysis to examine how criminal liability may arise when \GAI models are used to generate \CSAM. Because liability in such cases depends heavily on both legal definitions and specific technical circumstances, we adopted an interdisciplinary approach involving both legal experts and computer scientists.

The legal analysis centered on key German legal frameworks, including the German Criminal Code (\STGB), German Youth Protection Act, and relevant parts of the EU Artificial Intelligence Act (AI Act). It examined how liability could arise in different scenarios involving \GAI \CSAM. The assessment was led by two public prosecutors from the Bavarian Central Office for the Prosecution of Cybercrime, as well as two legal scholars specializing in the intersection of law and artificial intelligence.

\begin{figure}[ht!]
  \centering
  \includegraphics[width=0.9\columnwidth]{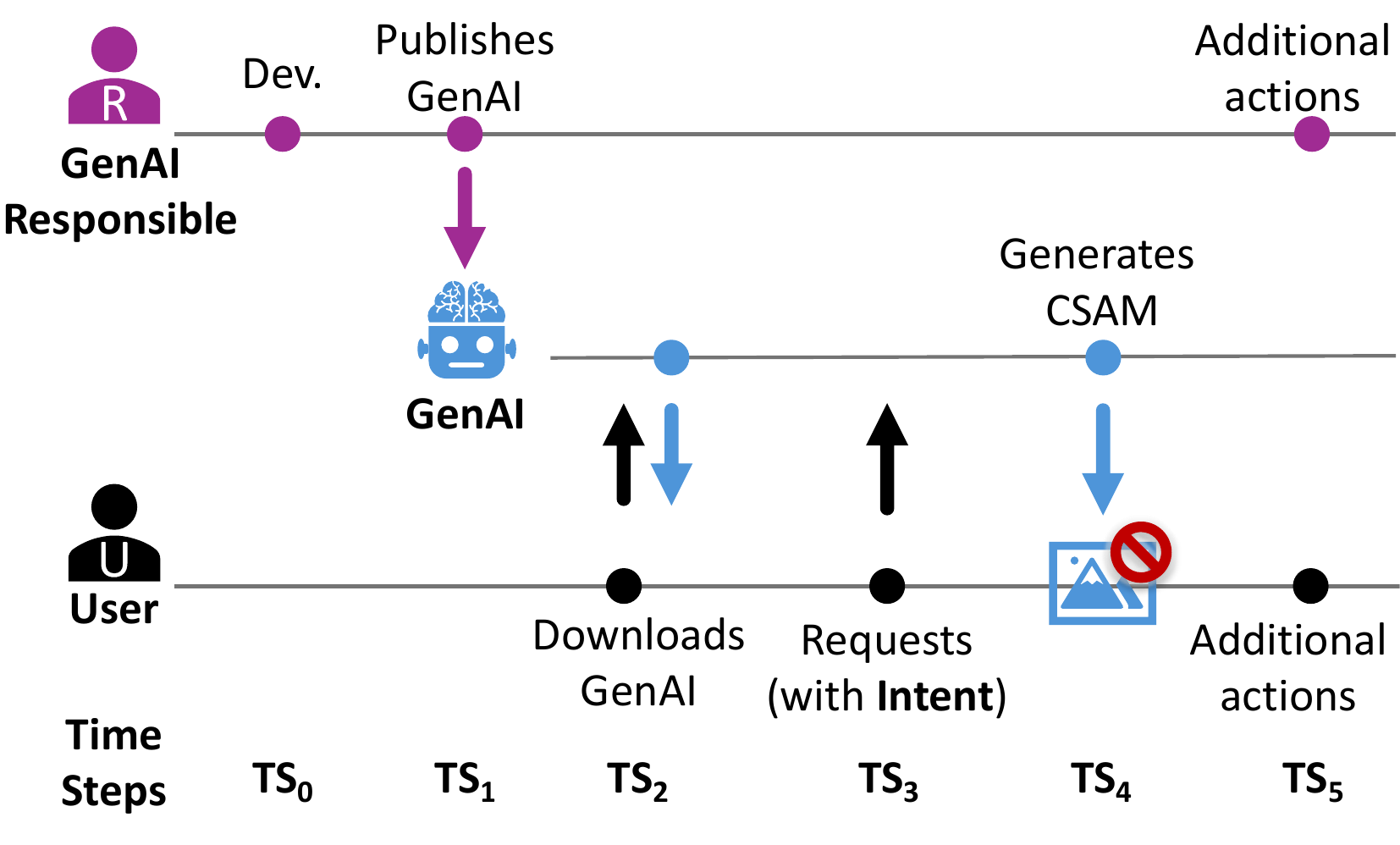} 
  \caption{Template scenario. It shows the three main actors and time steps of the generation of \CSAM imagery, including the \textit{main scenario} ($T_1-T_4$) and additional step considerations ($T_0$ Development and $T_5$ Additional actions).}
    \Description{Present the basic scenario. It is composed of multiple steps, from the development, the deployment, the usage of the model, and additional steps.}
  \label{fig:Basic_Scenario}
  \vspace{-0.5cm}
\end{figure}

To ground the legal interpretation with realistic technical contexts, a template scenario is developed by two computer scientists experienced in building AI-based software. This ensures that each case is both technically feasible and relevant. The template scenario (see \figref{fig:Basic_Scenario}) outlines the basic structure of the scenarios analyzed. Additionally, \tabref{tab:properties} describes technical and contextual, \GAI/system, properties that we can vary in the template scenario, \eg the type of the model ($P_1$) and purpose of the model ($P_2$). These features are selected due to their relevance for the legal analysis. The template scenario has the following structural elements: 

\begin{itemize}
    \item \textbf{Three main actors}:
    \begin{itemize}
        \item [\user] \textbf{User:} A natural person that uses the GenAI model to generate \CSAM. Unless specified otherwise, the \textit{User} is assumed to act intentionally, that is, with a deliberate aim of generating \CSAM.
    
        \item [\genAIR] \textbf{\GAI Responsible:} A natural person, who is the legal responsible of the GenAI model. This role may be filled by a researcher, the person legally responsible for a company (legal representative of a company), or an independent developer. Under German law, this is who may bear the responsibility for offenses committed or enabled by the model.

        \item [\model] \textbf{\GAI:} Generative artificial intelligence model used to generate the \CSAM content. 
    \end{itemize}
    \item  \textbf{Time steps with their associated actions}:
    \begin{itemize}
    \setlength\itemindent{0.5em}
        \item [$\bm{TS_0}$] \textbf{Model/System Development}. During this phase, the GenAI responsible defines the model's properties (see \tabref{tab:properties}).

        \item [$\bm{TS_1}$] \textbf{Model Publication.} The developed model/system is published. This can occur through direct releases (\eg on platforms such as Hugging Face or Civitai)  or as a hosted service/app (as defined in property $P_4$). $P_4$ may enable/facilitate additional properties, such as $P_5$ and $P_8-P_{11}$.
    
        \item [$\bm{TS_2}$] \textbf{Model Download.} If the model is a \texttt{standalone download}, this is the time step that marks \textit{user's access to the model}. Otherwise, access is considered to occur at $T_3$. Because not all \GAI models can be downloaded, this step is optional.
    
        \item[$\bm{TS_3}$] \textbf{\CSAM Request.} The \textit{User} \textit{intentionally} requests the generation of \CSAM content. 

        \item[$\bm{TS_4}$]  \textbf{\CSAM Generation.}  The \GAI model generates the requested \CSAM imagery and provides it to the user.
    
        \item[$\bm{TS_5}$] \textbf{Additional Actions.} The \CSAM content was generated, and additional actions can be executed, depending on the properties (\eg $P_9-P_{11}$) that the system/model has. For example,  the imagery can be shared through the developed system by the \textit{User}, or the generated \CSAM can be stored by the system.   
    \end{itemize}
    \vspace{-0.3em}
\end{itemize}

\begin{table*}[ht]
\centering
\caption{Technical and contextual properties of the \GAI model and their default values in the basic scenario. \textbf{Default} are shown in \texttt{\textbf{bold typewriter}} font. A \texttt{\textbf{not}} in that font indicates the default is the negation of the property.}
\label{tab:properties}
\begin{tabularx}{\textwidth}{>{\bfseries}p{0.3cm} >{\bfseries}p{5.5cm} X}
\toprule
\textbf{ID} & {\rmfamily \textbf{Title}} & \textbf{Description} \\
\midrule
$\bm{P_1}$ & Type of Model & \texttt{\textbf{Foundational model}} or modified-foundational model, \ie through additional training. \\
$\bm{P_2}$ & Purpose of the Model& \texttt{\textbf{General purpose}}, non-human related (\eg generation of catalog images of furniture), or human related purposes (\eg advertisement including humans)\\ 
$\bm{P_{2.1}}$ & Nudity & Nudity can be explicitly present or \texttt{\textbf{not}} in the purpose of the \GAI model. \\
$\bm{P_3}$ & Type of Imagery Generated & Artistic or \texttt{\textbf{realistic imagery}}. \\

$\bm{P_4}$ & How the Model is Published & The model can be a \texttt{\textbf{standalone download}}, be part of an app (\eg a chatbot), or offered as a service (\eg Rest API). \\
$\bm{P_5}$ & Geographical Aspect & It involves the location of the server, the location of the \textit{User}, or the nationality of the \textit{User}. We distinguish between within \texttt{\textbf{Germany}} or outside it.\\
$\bm{P_6}$ & \acrfull{tou}& The restriction of  not using the model for \CSAM generation is explicitly mentioned or \texttt{\textbf{not}} in terms of use associated to the model. \\
$\bm{P_7}$ & Content Moderation & The content moderation of unlawful material generated by the \GAI in relation to \CSAM can be state-of-the-art, \texttt{\textbf{non state-of-the-art}}, or nonexistent. \\
$\bm{P_8}$ & Access to Internet & The \GAI app or service is reachable via Internet or \texttt{\textbf{not}}. \\

$\bm{P_9}$ & Storage of User Requests and Images & The generated imagery and requests can be stored or \texttt{\textbf{not}}. \\
$\bm{P_{10}}$ & Share Results of Prompts & Apps or services can offer the functionality to share prompts and generated results (incl. \CSAM imagery) or \texttt{\textbf{not}}. \\
$\bm{P_{11}}$ & Monetize with the Users' Histories & Prompts and generated results (incl. possibly \CSAM imagery) can be used for monetization (\eg selling of data) or \texttt{\textbf{not}}. \\
\bottomrule
\end{tabularx}
\end{table*}

From this template, we create our \textbf{baseline scenario} that serves as the starting point for our statuary interpretation, which uses the default values for all the properties in \tabref{tab:properties}. These combinations give us the following:



\begin{tcolorbox}[colback=myblue!1,colframe=myblue!50!black,boxrule=0.5pt, breakable, top=1pt, bottom=1pt, boxsep=1pt, before skip=5pt, after skip=5pt]
\textbf{Baseline Scenario:} A \textit{User} requests the generation of \CSAM ($\text{TS}_3$),  and it is  being generated ($\text{TS}_4$) by a \texttt{foundational} ($P_1$), \texttt{general-purpose} \GAI model ($P_2$). The model is targeted to create \texttt{realistic} imagery ($P_3$) and has been published as a \texttt{standalone download}  ($P_4$) to the public rather than deployed as part of a hosted service or application ($\text{TS}_1$).  The \textit{User}, located in \texttt{Germany} ($P_5$), accesses the model by  downloading it locally ($\text{TS}_2$).  The terms of use ($P_6$) associated with the \GAI \texttt{do not explicitly prohibit} \CSAM-related prompts or outputs.  During training ($\text{TS}_0$), the \textit{Responsible} applied \texttt{state-of-the-art} content moderation practices ($P_7$) to reduce the likelihood of \CSAM generation, but the model itself does not retain any connection to moderation infrastructure after release. This is also strengthened by the fact that the model is  \texttt{not reachable via the internet}, limiting the control and oversight from the \textit{Responsible}. Additionally, since just the \GAI model is released\textemdash not an app\textemdash the \GAI \textit{Responsible} does not have control over the data generated\textemdash they cannot store the requests or generated data ($P_9$), nor actively enable the sharing of it ($P_{10}$) or monetizing it ($P_{11}$).
\end{tcolorbox}



We then develop a series of modified scenarios, each altering one or more parameters from the baseline to reflect different contextual or technical properties. Examples of such modifications include using a \texttt{modified-foundation model} ($P_1$), \texttt{deploying the model within an app} ($P_4$), or \texttt{having internet access} ($P_5$).

The scenarios are independently analyzed by the public prosecutors and the legal scholars. This separation means we first obtain two independent perspectives, both created by experts from different backgrounds: public prosecutors, who are involved in the application and enforcement of laws, and legal scholars, who rather interpret laws. Each group of experts applies \textit{statutory interpretation} methods to assess whether, and how, criminal liability could arise in the given circumstances. After the independent assessments, the responses are reviewed by the computer scientists, who compare interpretations and identify differences and common patterns.

The outcome of the merged results is validated by the legal experts. And finally, any discrepancies between the experts’ interpretations are discussed in joint follow-up sessions involving both legal and computer scientists. The aim is to resolve ambiguities and ensure that legal reasoning accurately reflects the technical realities of each scenario.

\section{Results}
\label{sec::results}

This section presents the key outcomes from the \textit{statutory interpretation} and the joint analysis process. In order to avoid redundancy, the results of the legal analysis\textemdash specifically, who may be held criminally liable and which factors (\ie properties) influence that assessment\textemdash are presented incrementally. This means that we begin with the general analysis of the \textit{baseline scenario}, followed by an examination of individual property variations that have a legal impact on liability. As a guiding framework throughout this analysis, unless otherwise stated, the \textit{statutory interpretation} is based on \textit{\law~\textit{184b}~\STGB}~\textit{(Dissemination, procurement and possession of child pornographic content)}. Note that while the initial interpretations of the experts were not equal, the subsequent discussion successfully resolved all discrepancies, such that these results reflect the opinions of all legal experts involved.

\subsection{Analysis for the Basic Scenario}

In these cases where realistic \CSAM imagery is generated using a \GAI model, two main offenses can be identified. The \textbf{primary offense} is the \textit{act of generating \CSAM itself} (see \figref{fig:german_law}, 8). Additionally, a \textbf{secondary offense} of \textit{aiding and abetting} may be considered (see \figref{fig:german_law}, 7) in these scenarios. 

With regard to the criminal liability, the main offender\textemdash or \textbf{perpetrator}\textemdash will be the \textit{User}, as they are the one who actively initiates the generation of \CSAM using the \GAI 
(\figref{fig:Basic_Scenario}, $\text{TS}_3$). The \GAI \textit{Responsible} may be considered a secondary party or \textbf{participator} to the offense, potentially bearing liability for aiding and abetting by providing access to a model capable of producing such content. 

For both the principal and secondary offenses, it is necessary to (I) prove the \textit{offender}'s intent (\ie \textit{User's} intent or \GAI \textit{Responsible's intent}) and/or (II) their knowledge of the likely consequences of their actions. This means the offenses are committed \textit{willingly} and/or \textit{knowingly}. It is precisely at this point that variations in the studied properties (see \tabref{tab:properties}) influence the assessment of criminal liability of the \GAI \textit{Responsible}\textemdash either aggravating or mitigating the likelihood of establishing these elements.

Note that, in the context of \CSAM, \textit{negligence} is not a relevant category since this offense by law can only be committed intentionally\textemdash\ie meaning that it is always treated as an \textit{intentional} act\textemdash as opposed to, for example, causing bodily harm (see Section \ref{sec::gLAW}).
\vspace{-0.2em}
\subsection{Variations of the Basic Scenario}
In the following subsections, we will analyze how different variations of the properties may affect the assessment of criminal liability of our \textbf{Basic Scenario}. Particularly, we present the analysis divided into two parts: variations primarily relevant from $\text{TS}_0$ (Development) through $\text{TS}_4$ (\CSAM Generation); and variations relevant to the final time step, \ie $\text{TS}_5$ (Additional actions).

\subsubsection{Important Properties Variations Between Time Steps $\text{TS}_0$ (Development) and $\text{TS}_4$ (\CSAM Generation)}

\mypara{\textbf{$\bm{P_1}$ Type of Model}}
When a \texttt{modified foundational model} is used, two situations arise regarding who is responsible for the actions enabled by the \GAI. In the first situation, the model is modified by the original developer of the \texttt{foundational model}. The second situation is when a third party modifies the \GAI\textemdash a party that did not develop the foundational model.  

In both situations, the primary and secondary offenses remain the same. What changes is who is held criminally liable for the secondary offense. This is because the contributions of each of the parties involved in developing the \GAI must be considered, as well as how those contributions relate to the offense of aiding and abetting the generation of \CSAM.  Thus, in the first situation, where the \GAI original developer modifies the model, the \GAI \textit{Responsible} remains the same, regardless of when the modifications enabling the \CSAM generation were introduced \textemdash in the foundational or modified model. In the second situation, however,  criminal liability primarily shifts to whoever last modified the \GAI. This is because liability for the developer of the foundational model is more difficult to establish, as they are further removed from the primary perpetrator, \ie the \textit{User}.

\mypara{\textbf{$\bm{P_2}$  Purpose of the Model}} In this case, regardless of the purpose of the \GAI model\textemdash \ie whether it is general purpose, explicitly mentions the generations of non-human-related imagery, or explicitly mentions the generation of imagery related to humans\textemdash neither the possible offenses nor the criminal liabilities associated with them are affected.  There are exceptions for this, which are covered in the analysis of $P_{2.1}$ \texttt{Nudity}.

\mypara{\textbf{$\bm{P_{2.1}}$ Nudity}} When \texttt{nudity} is explicitly mentioned in a model’s stated purpose as a possible type of content, that mention can help to prove intent. In particular, when circumstances suggest misuse, it would be relevant to establish \textit{intent} in the secondary offense of aiding and abetting. The extreme case is one in which \CSAM is or is part of the model's purpose\textemdash indeed, this is the case for some LoRAs, especially those found on the Darknet (\ie fine-tuned image generation models with Low-Rank Adaptation)~\cite{IWF2024AI_CSAM, Schurig2025Pulitzer}. In that case, the intention of supporting the act of generating \CSAM will be clear.

\mypara{\textbf{$\bm{P_{3}}$ Type of Imagery Generated}}  When generated imagery is not \texttt{realistic}, the mentioned criminal offenses may change or be nonexistent. In cases where the generated  \CSAM imagery is \textit{artistic} (see Section \ref{sec::background}), and the \textit{User} has no intention to disseminate it, \ie making the content publicly available or making it available to a third party that makes it publicly available, no criminal offense is present under the German law. However, because  \GAI \textit{Responsible} generally cannot determine users' intentions, the ability to generate such content remains a legal risk. Therefore, limiting the \GAI just to generate \texttt{artistic} imaginary does not necessarily eliminate the criminal liability.  

\mypara{\textbf{$\bm{P_{4}}$  How the Model is Published}} If the model is not released as a  \texttt{standalone download}  but is instead integrated into an  \texttt{app} or as a \texttt{service}, this can have implications in three areas: (I) the new capabilities and functionalities enabled in the system; (II)  the potential knowledge about  \CSAM generation; and (III) the assignment of responsibility in the secondary offense. 

Regarding the first, the potential implications are discussed under other properties because they are the enabled functionalities or capabilities.  In particular,  how the content moderation is applied ($P_{7}$),  the model having  access to internet ($P_8$),  storing, sharing and monetizing user history (\ie $P_{9}$, $P_{10}$, $P_{11}$ respectively).

In terms of the second\textemdash potential knowledge of \CSAM generation\textemdash integration into a service may allow the \GAI  \textit{Responsible} to gain insight into how the model is used and what type of content is being generated. This could increase the \textit{Responsible's} awareness of  \CSAM-related misuse and potentially trigger legal obligations (\eg reporting of offenders to the authorities, restricting their access, and/or taking preventive measures).  

When analyzing the third implication\textemdash namely the assignment of liability for the secondary offense\textemdash this follows a line of reasoning similar to that discussed under ${P_1}$ Type of Model. The main aspect to consider will be that the model is released by a third party distinct from the original development party. In such cases, the respective contributions of the developing party and the releasing party must be analyzed for the secondary offense of aiding and abetting. 

\mypara{\textbf{$\bm{P_{5}}$ Geographical Aspect}} For this property, it is necessary to distinguish (I) the territory in which the offenses are carried out from (II) the nationality of the actors.

Regarding the location of the offenses, we analyze two places: the location of the primary offense and the location of the secondary offense. Under \textit{\law~\textit{9}~\STGB}, the primary offense is situated where the perpetrator\textemdash in our case, the \textit{User}\textemdash acts. The secondary offense, according to the same rule, takes place both at the location of the main offense and in any other location where the participator \textemdash in our case, the \GAI \textit{Responsible}\textemdash acts. This means that, even if the server is \texttt{not located in Germany}, the offenses can still be considered under German law if the perpetrator or participator is within German territory.

As to nationality, the status of the \textit{User} and the \GAI \textit{Responsible}, as German nationals or non-nationals, influences the applicability of German criminal law when the conduct occurs \texttt{outside German territory}. Specifically,  \textit{\law~\textit{7}~\STGB} enables the two offenses to be prosecuted under German law if the \GAI service is provided by a \texttt{German national} or is used by a \texttt{German national} and the offense is also punishable at the place where it is committed.

\CSAM also falls under an even broader paragraph of the German criminal law. \law~\textit{6}~\STGB even allows the prosecution of certain offenses in Germany, even if no German nationals are involved and the offense happened outside of Germany, for crimes against internationally protected legal interests\textemdash which includes \CSAM. Consequently, German authorities can possibly prosecute any \GAI Responsible for \CSAM-related offenses (incl. offenses discussed in (see $\bm{P_{10}},\bm{P_{11}}$)), even if it happens elsewhere.


\mypara{\textbf{$\bm{P_{6}}$ \acrlong{tou}}}   The Terms of Use are a civil law contract between the \GAI \textit{Responsible} and the \textit{User}; their breach has only civil law consequences and therefore falls outside the criminal-law scope of this study. However,  an \texttt{explicit clause forbidding the generation of} \CSAM does not shield the \GAI \textit{Responsible} from criminal liability for aiding and abetting if \CSAM is generated.  On the contrary, the  \texttt{explicit}  mention can serve as evidence that the operator foresaw\textemdash and accepted!\textemdash the risk of \CSAM generation. This might be used as proof of intent, specifically the acknowledgment of the risk, required for aiding and abetting.

\mypara{\textbf{$\bm{P_{7}}$ Content Moderation}}  When analyzing how the changes in the level of content moderation affect the analysis of our basic scenario, we identified that a key aspect is whether the \GAI \textit{Responsible} has implemented  \texttt{state-of-the-art} content moderation safeguards. When such measurements reduce the risk of misuse of the \GAI \textemdash generating \CSAM\textemdash to a level still tolerable under  criminal law, the offense of aiding and abetting is unlikely to be considered.  In contrast, if moderation is \texttt{nonexistent} or is \texttt{non-state-of-the-art}, this may indicate a conscious acceptance of the risk of \CSAM  creation. That acceptance may satisfy the intent element of aiding and abetting, making criminal liability possible.

Another key question is when content moderation should be in place. \texttt{State-of-the-art} measurements must exist at the \textit{moment the User gains technical access to the model},  not merely when the request for \CSAM is made.  This ``\textit{moment of access}'' depends on the publication method described in $P_4$. If the model is released as a \texttt{standalone download} and the \GAI \textit{Responsible} does not have control over it, the decisive moment is the download itself (\figref{fig:Basic_Scenario}, $\text{TS}_2$). Otherwise, it is assumed that \textit{Responsible} has access to it and the model can be updated, then the relevant moment recurs every time the \textit{User} requests (see \figref{fig:Basic_Scenario}, $\text{TS}_3$).

\mypara{\textbf{$\bm{P_{8}}$ Access to Internet}}  When a model remains reachable \texttt{via the Internet},  the \GAI \textit{Responsible} may retain continuous technical control over the model.  This implies that the \GAI \textit{Responsible} can now implement an oversight mechanism to monitor the model, possibly increasing the awareness of \CSAM generation (
discussed in $P_4$),  as well as enabling continuous \textit{content moderation} and update policies. If needed to prevent illegal use, this also allows providers to completely prevent future use of their services, \ie suspend or shut down the service altogether.  All of this contrasts with the offline version\textemdash model not reachable via the internet\textemdash  where the awareness of the usage and corrective actions is limited.

\subsubsection{Important Properties Variations in Time Step $\text{TS}_5$ (Additional Actions)} 

In this section, we assume that \CSAM was already generated  (\figref{fig:Basic_Scenario}, $\text{TS}_5$). Consequently, the two offenses discussed in the \textit{basic scenario} \textemdash \CSAM generation and aiding and abetting this primary act\textemdash are no longer the only legal concerns.  Three capabilities that were inactive in the \textit{basic scenario}\textemdash storage, sharing, and monetization of requests and generated data\textemdash now come into play and may lead to \textit{additional} offenses.

\mypara{\textbf{$\bm{P_{9}}$ Storage of User Requests and Images}}  If the \textit{User's} requests and data generated containing \CSAM are being stored, criminal liability does not arise automatically for the \GAI \textit{Responsible}. Under German law\textemdash \law \textit{\textit{7 para. 1}} of the \textit{\textit{German Digital Services Act} } joint with Article \textit{\textit{6 para.1 (EU) 2022/2065}} \textemdash storing illegal content\textemdash \CSAM in our scenario\textemdash  will not lead to criminal liability for the \GAI \textit{Responsible} as long as  (I) they do not gain actual knowledge of it and (II) once they do,  they must act without delay to block or remove the material. Failing to do so means that knowingly \texttt{storing} \CSAM  may be viewed as an intentional attempt to enable the storage of incriminating content, which may trigger criminal liability to the \GAI Responsible.

\mypara{\textbf{$\bm{P_{10}}$  Share Results of Prompts}}  The same key aspect\textemdash having knowledge of the act, sharing \CSAM{}\textemdash governs the sharing of results and requests.  If the \GAI \textit{Responsible} is aware that \CSAM is shared through the service and nonetheless allows it, the main offense may be the dissemination of \CSAM and the secondary offense would be aiding and abetting the distribution of such content. This consideration depends on how the sharing is supported, \eg if prompts or outcomes are shared directly by the \GAI Provider or if the sharing is only possible by providing others with links. 




\mypara{\textbf{$\bm{P_{11}}$ Monetize with the Users' Histories}} If the  \GAI \textit{Responsible} wants to monetize  \textit{User's} history, for example, by selling the requests and generated data, it may lead to additional criminal liability. As with the previous two properties, \GAI \textit{Responsible's} awareness of the existence of \CSAM in the data is a key aspect.  If they are aware of it, they will not only support the retrieval of \CSAM, but they could also be charged as an independent, primary offender for distributing  \CSAM.

\subsection{Additional Considerations}

Our analysis so far already shows that there are many factors that influence the potential legal exposure of developers and companies. To conclude our legal analysis, we go beyond the pure application of the relevant aspects from criminal law for \CSAM to shed light on other important legal aspects that interact with this.

\subsubsection{Intentionality} One aspect that may affect the determination of the existence of the intention from the \GAI \textit{Responsible} in the offense of aiding and abetting is their technical expertise.   This means that they could rely on the fact that no \CSAM  content will be produced or that nobody will generate it. However,  given the typical use of \GAI,  one could argue that technically experienced users do not fully trust that there will be no misuse and therefore act with intent, \eg \cite{thiel2023generative, thiel2023U, iwf2023Generation}.

Regulations and other laws can also be a factor with respect to intent. For example, compliance with the \textit{European Union's AI Act} requires several activities on the part of model providers. For example, any system that is considered high risk (\textit{Art. 6}) must implement a risk management system (\textit{Art. 9}), including post-market monitoring proportionate to the risk (\textit{Art. 72}). The AI Act also regulates that providers of general-purpose AI models with systemic risk (\textit{Art. 51})\textemdash very large models, including Chatbots with image generation capabilities\textemdash have similar obligations for risk assessment, mitigation, and monitoring (\textit{Art. 55}). Arguably, the risk management should identify the possibility of generating \CSAM content. In addition, the post-market monitoring should also detect if this happens, possibly leading to further obligations to implement measures to prevent \CSAM generation.

This shows how compliance\textemdash or non-compliance\textemdash with market laws and regulations possibly affects criminal liability, because it can be a factor in determining intent for criminal acts, \eg aiding and abetting \CSAM generation. 

\subsubsection{Criminal Trading Platform}

A fairly recent addition to Germany's criminal code is \textit{\law~127~\STGB (Operating criminal trading platforms on the internet)}. Depending on the interpretation of the role played by a platform that hosts a \GAI that has as its purpose \CSAM generation, where users can obtain and share such content, it may be treated as a criminal trading platform or not. However, since the law is relatively new, the exact scope of the law remains unclear, as courts have not yet clarified its precise boundaries.

\subsubsection{Protection of Minors}

Another aspect is that minors may also use the \GAI when it is released to the general public (\figref{fig:Basic_Scenario}, $\text{TS}_1$). This means that the \GAI providers may also be criminally liable under \textit{\law 27 JuSchG}  in conjunction with  \textit{\law 15 JuSchG para. 2 no. 1}~\cite{JuSchG2024} for making harmful media available to minors. The key question that needs to be evaluated is whether publishing the model is actually making it accessible to minors by default. Furthermore, if this is the case, this would not only imply an obligation to prevent \CSAM, but also other harmful content, including, \eg extreme violence or pornographic content that does not fall under \CSAM. 

\section{Discussion}
\label{sec::discussion}

Our analysis revealed that criminal liability for those responsible of the \GAI when \CSAM  is generated is affected by technical and contextual features related to the \GAI, and in some cases, even change the type of liability (\eg from secondary to primary). These variables include whether the model has been modified, how it has been published, where it is executed, and the capabilities of the model or the system in which it is located. Specifically, the variation on these aspects will help to prove intention behind the actions of  the \textit{User} and the \GAI \textit{Responsible}.  This has implications not only for those who are viewed as the responsible party for the model, but also for those who contribute to the development of \GAI. 


\begin{tcolorbox}[colback=myblue!1,colframe=myblue!50!black,boxrule=0.5pt, breakable, left=2pt, right=2pt, top=1pt, bottom=1pt, boxsep=1pt, before skip=5pt, after skip=5pt, title=Highlight 1]
Parties \textit{responsible} for \GAI may primarily face \textbf{secondary} liability; however, \textbf{primary} liability is also possible if specific  conditions are met, \eg enabling sharing of generated content via the platform.
\end{tcolorbox} 

\vspace{-0.5em}
\subsection{Implications for Developers and GenAI providers}


Across the end-to-end \GAI life-cycle, and in any system that incorporates \GAI there are actions that can be considered to prevent criminal misuse.  First, developers should ensure they develop a system with \texttt{state-of-the-art} content moderation measures against \CSAM generation.  These measures begin with understanding the training data and ensuring that such illegal content is not present \cite{nist2024synthetic, nist2024comments, thorn2024safety, unicri2024generativeai, thorn2024solutions}. Post-training to prevent models from acting on harmful inputs~\cite{dai2023safe,piet2024jatmo} and make prompt-hacking, \eg red teaming~\cite{nist2024comments, nist2024synthetic,thorn2024solutions}, are additional content moderation measures that should be used. Post-deployment, active content moderation by monitoring system usage should also be considered. Additionally, since the time of access is relevant to criminal liability, updates of content moderation techniques should be a normal part of the model maintenance.



\begin{tcolorbox}[colback=myblue!1,colframe=myblue!50!black,boxrule=0.5pt, breakable, left=2pt, right=2pt, top=1pt, bottom=1pt, boxsep=1pt, before skip=5pt, after skip=5pt, title=Highlight 2]
Define and enforce \CSAM \textbf{timely} content-moderation policies \textbf{across} the \GAI \textbf{life-cycle}\textemdash such as clean training data, red-teaming, active monitoring, and continuous updates\textemdash is a 
priority requirement to \textbf{mitigate criminal misuse and liability risk}.
\end{tcolorbox} 

The ability to moderate GenAI after deployment leads to another implication: \texttt{models hosted} by providers or \texttt{integrated into apps}  reachable via \texttt{internet} should be preferred over \texttt{standalone, downloadable models}. On the one hand, this allows the \GAI provider to learn about the generation/dissemination of \CSAM. However,  this knowledge may already exist due to external factors and technical expertise regarding possible misuse. On the other hand, this will facilitate taking action if needed. These actions include updating content moderation policies, removing stored illegal content, and possibly even removing the models if a model is proven capable of generating \CSAM and preventive measures are ineffective. 

\begin{tcolorbox}[colback=myblue!1,colframe=myblue!50!black,boxrule=0.5pt, breakable, left=2pt, right=2pt, top=1pt, bottom=1pt, boxsep=1pt, before skip=5pt, after skip=5pt, title=Highlight 3]
Prefer provider-\textbf{hosted} \GAI, because centralized control enables post-deployment moderation and quick actions against misuse.
\end{tcolorbox} 


For the \textbf{\GAI providers}\textemdash who, from a legal standpoint, is the \textit{Responsible} party of the model, who can be a \textit{researcher}, independent \textit{developer}, or a \textit{company responsible}\textemdash there are also broad, less-technical measures \textit{to consider alongside developing focused suggestions}.  First, even if the model does not generate \texttt{realistic} imagery, and the \TOU \texttt{expressly} forbid \CSAM generation, no criminal safe harbor applies: liability may still arise if recognizable \texttt{artistic} \CSAM is produced with the intention of dissemination, and the \texttt{explicit} prohibition is present, due to this prohibition can itself demonstrate that the provider foresaw the risk.  Moreover, achieving this safe harbor is not possible regardless of when the provider participated in the development or where the model is deployed. Thus, even if a third party releases a \texttt{modified} model, criminal liability may be assigned to any contributor whose actions enabled the misuse. Additionally, the German Criminal Code allows the prosecution of any offense that was partially committed on German territory or by a German national, as well as offenses committed entirely outside Germany by non-Germans. Therefore, not even hosting the model outside of Germany and offering it only in other countries is a guarantee of German criminal safe harbor.

\begin{tcolorbox}[colback=myblue!1,colframe=myblue!50!black,boxrule=0.5pt, breakable, left=2pt, right=2pt, top=1pt, bottom=1pt, boxsep=1pt, before skip=5pt, after skip=5pt, title=Highlight 4]
\GAI \textbf{providers} and \textbf{contributors} \textbf{cannot} rely on \textbf{\TOU bans} or \textbf{foreign hosting} for immunity: they may still face criminal liability\textemdash including under Germany’s broad jurisdiction\textemdash even where \textbf{non-photorealistic} \CSAM is generated and disseminated.
\end{tcolorbox} 


Additionally, the \GAI provider should \textit{promptly implement measures} against the misuse once it becomes aware of \CSAM in the generated data. Failure to do so may increase the provider's potential criminal liability. These measures include removing any detected \CSAM, updating content-moderation policies, or shutting down the service if a model is known to be capable of generating \CSAM \cite{nist2024synthetic, nist2024comments, thorn2024safety}. This is particularly important when the data containing \CSAM is \texttt{stored}, \texttt{shareable} through the system or \texttt{monetized}.  The importance lies in prompt measures needed to avoid potential additional offenses, such as being the perpetrator of distributing \CSAM. Deciding against such measures due to possibly high costs could be used as evidence of an active act. 

\begin{tcolorbox}[colback=myblue!1,colframe=myblue!50!black,boxrule=0.5pt, breakable, left=2pt, right=2pt, top=1pt, bottom=1pt, boxsep=1pt, before skip=5pt, after skip=5pt, title=Highlight 5]
Upon \textbf{becoming aware} of \CSAM in generated data, providers must act \textbf{promptly}, remove detected \CSAM, update moderation, or shut down the service, \textbf{to avoid increased} criminal liability, especially when the \CSAM is \textbf{stored}, \textbf{shareable}, or \textbf{monetized}.
\end{tcolorbox} 

Furthermore, if a model has the purpose of generating \texttt{nudity}, it should be accompanied by a thorough evaluation of the risk of \CSAM generation and strong countermeasures. This is because the purpose of generating \texttt{nudity} itself does not strongly impact determining criminal liability. However, if the \GAI  explicitly mentions the purpose of generating \CSAM, it shows intent; or if the model has a more general purpose of generating \texttt{nudity}, it will help to prove intent of supporting \CSAM generation. 

\begin{tcolorbox}[colback=myblue!1,colframe=myblue!50!black,boxrule=0.5pt, breakable, left=2pt, right=2pt, top=2pt, bottom=2pt, boxsep=1pt, before skip=6pt, after skip=6pt, title=Highlight 6]
\textbf{Nudity purpose models} must include \textbf{strong} \CSAM-risk evaluation and safeguards; \textbf{explicit \CSAM} purpose shows intent, and even a general \textbf{nudity} purpose \textbf{may support} it.
\end{tcolorbox} 

\subsection{Implications for Researchers}

Researchers who work on \GAI also bear responsibility for the models they publish. Consequently, \textit{the previous suggestions and advice also apply to researchers}, but additional considerations deserve special attention, particularly when publishing open-source models\textemdash whether as standalone releases or part of a replication package. First, in an ideal scenario, the model should be designed and released in a manner that allows for ongoing monitoring, rather than simply being deployed and then forgotten, especially if not only the weights are published but the model is also offered as a service for demonstration purposes. Second, if models are stored in long-term archives, it has to be considered that their immutability and durable storage may prevent corrective or preventive measures once misuse is detected.  Both points are especially important because content moderation policies need to be \texttt{state-of-the-art} (SOTA) at the moment users first gain access\textemdash that is, at download time, not at publication.  Third, \textit{researcher} should recognize that by releasing a model, they enable third parties to modify and republish it, and the original developer, the researcher, may still face criminal liability if the model is later misused. 

\begin{tcolorbox}[colback=myblue!1,colframe=myblue!50!black,boxrule=0.5pt, breakable, left=2pt, right=2pt, top=1pt, bottom=1pt, boxsep=1pt, before skip=5pt, after skip=5pt, title=Highlight 7]
\textbf{Researchers} are responsible for what they release, even if \textbf{others later modify} it. Ensuring \textbf{SOTA} content-moderation safeguards at \textbf{first user access} is recommended, implying \textbf{continuous} updates and\textbf{ maintained} access for \textbf{modification}.
\end{tcolorbox} 

This implies that, to support open science while reducing the risk of criminal liability and given limited post-release control, the publication of \GAI image models\textemdash especially photorealistic ones\textemdash requires stricter protocols. These include stronger data-cleaning policies, staged releases that allow for controlled and rigorous testing, and long-term archiving removal policies (\eg \cite{internetarchive, zenodo}) that facilitate the removal of resources in these special cases.

\subsection{Implications for Policymakers}

Our analysis also raises an important question that policy makers need to answer and that companies, lobbyists, and other affected parties can influence: \textit{Is the current legal situation good as it is or are changes to laws and regulations required? }

Policymakers might conclude that the current regulations are too weak because they do not explicitly require \CSAM moderation and prevention by \GAI model providers. We derive such requirements indirectly by considering whether this failure to implement preventive measures could be an intentional act that aids and abets criminals. In such a case, policymakers might modify or extend laws to make the obligation clear and directly regulate the (criminal or civil) consequences of failing to act with respect to \CSAM prevention. 

However, policymakers might also conclude that the current regulations might hinder the development of AI products and research. For example, sharing open-weight models by researchers may be unintentionally criminalized, or start-ups that simply re-use or fine-tune models would have to implement\textemdash possibly extensive and expensive\textemdash content moderation measures. In such a case, policymakers might modify or extend laws to clarify which uses would require less moderation. 

\begin{tcolorbox}[colback=myblue!1,colframe=myblue!50!black,boxrule=0.5pt, breakable, left=2pt, right=2pt, top=1pt, bottom=1pt, boxsep=1pt, before skip=5pt, after skip=5pt, title=Highlight 8]
We require \textbf{clearer} \CSAM legislation for \GAI \textbf{providers} and define \textbf{consequences} for failures. They should consider \textbf{low-risk uses to protect innovation}.
\end{tcolorbox} 
\subsection{Realism of Criminal Prosecution}

Now that we have considered many aspects, let us return to our initial hypothetical: Is it actually realistic that the CEO of a \GAI company, a researchers, or a software developer is arrested, because their products are used to generate \CSAM? 

There are certainly cases in which prosecution is more likely, especially if models or products are created for use cases that include pornography. One may argue that this is a gray market anyway, and people producing products for such gray markets are aware of the risks involved. 

However, that new technology\textemdash in our case \GAI{}\textemdash can also lead to the criminal prosecution of seemingly harmless products is nothing new. Specifically, in the context of \CSAM, this had occurred in Germany in 1998, when the internet was still establishing itself as a new, broadly available technology. At that time, a German district court actually sentenced the CEO of the German dependency of the internet provider CompuServe to two years in jail (on probation), because users were accessing \CSAM content.\footnote{AG M\"unchen, 28.05.1998 - 8340 Ds 465 Js 173158/95} While this ruling was later overruled by the state court,\footnote{
LG M\"unchen I, 17.11.1999 - 20 Ns 465 Js 173158/95} it took until 2007 to create legal certainty for providers in Germany with respect to criminal content that is accessed by their users, through new laws \textit{(Telemediengesetz (TMG))}. Thus, we can only hope that such a situation does not arise for GenAI developers, providers, or researchers, and that our analysis helps responsible actors clarify how they will be regulated and assessed.

\begin{tcolorbox}[colback=myblue!1,colframe=myblue!50!black,boxrule=0.5pt, breakable, left=2pt, right=2pt, top=2pt, bottom=2pt, boxsep=1pt, before skip=5pt, after skip=5pt, title=Highlight 9]
Whether prosecutions of \GAI providers, developers, or researchers are realistic remains unclear\textemdash an ambiguity that calls for caution and robust safeguards.
\end{tcolorbox} 
\section{Limitations}
\label{sec::limitations}

Our study has a few limitations that are relevant to understanding our results. The first one relates to its \textit{geographical scope}. Although we based our analysis on one of the major global players, which has a strong legal framework and is strongly influenced by a broader legal framework (\ie the EU legal framework), legal frameworks vary across jurisdictions (\eg artistic imagery, such as \CSAM in Manga, are not a criminal offense in Japan~\cite{kokolaki2025unveiling}).

The second limitation concerns the \textit{interpretation of the law}. While our analysis reflects the perspective of multiple legal experts\textemdash two prosecutors and two legal scholars specializing in the intersection of law and artificial intelligence\textemdash other experts on German law might disagree. Then, our analysis prescribes only a possible future risk, which would depend on multiple circumstances occurring together: the prosecution of \GAI-generated  \CSAM  case in Germany, a prosecutor who interprets the providers' actions as aiding and abetting, and judges who agree with this notion and proceed to convict. To the best of our knowledge, there are no pending criminal proceedings against a \GAI provider.

\section{Conclusion}
\label{sec::conclusion}

This multi-disciplinary study contributes to increasing awareness of the possible existence of criminal liability when \CSAM is generated by \GAI under German law. This is achieved by conducting \textit{statutory interpretation} alongside a series of scenarios in which we varied key properties influencing the existence of liability. 

Our findings showed that there may be different possible criminal offenses when dealing with a \GAI that generates \CSAM. The main offense is the \CSAM generation, in which the \textit{User} is the offender. \GAI \textit{Responsibles} might be secondary offenders by supporting the generation of this content through their technology. Furthermore, there are additional offenses such as distributing \CSAM that can be committed by the \textit{User} and the \GAI \textit{Responsible} as main offenders.

In addition to the possible identified offense, we have also identified that the criminal liability primarily depends on the intention behind the action and the knowledge of its consequences. Moreover, the variation of some properties\textemdash such as \texttt{not having state-of-the-art} content moderation measures at the moment when the \textit{User} has access to it\textemdash may increase the likelihood of establishing intent for the secondary offenders, \ie the \GAI \textit{responsibles}. Furthermore, some property variations that may create the idea of a legal safe harbor\textemdash such as the generation of just \texttt{artistic} imagery or hosting models in other countries\textemdash do not actually guarantee it. 

Based on those findings, we have discussed implications for a possible \GAI \textit{Responsible}\textemdash developers, researchers, and the person legally responsible for the company\textemdash or a person just fulfilling a role under a company that is not responsible for the model. There, we have identified possible actions to mitigate the risk of being criminally liable that can be considered by different actors, \ie developers, researchers, and legal representatives of a company.  

In light of this study and the inherent complexity of multidisciplinary studies\textemdash such as examining the implications of the gray areas surrounding the technical advancements, such as AI-generated \CSAM for people involved in their development and provision\textemdash it is recommended that further research be conducted. In particular,  studies that broaden the scope by exploring different legislation\textemdash not German legislation\textemdash on the same topic, \ie \CSAM, or by understanding different gray areas, in the same or different legislation, are especially encouraged. These will contribute to making our software engineering community even more aware of the possible legal implications when dealing with and developing such technology, \GAI. Furthermore, clear guidelines and possibly even legally binding regulations and standards that define requirements on how to mitigate the risk of \CSAM generation to provide legal safety for developers should be developed. While this is a challenging task due to the fast-paced development of \GAI technology, the severity of the misuse potential as well as the legal exposure that a \GAI responsible currently has, still mandates that such research and policy development take place.

\bibliographystyle{ACM-Reference-Format}
\bibliography{software}


\end{document}